\documentclass[usenatbib,usegraphicx]{mn2e}
\usepackage{float}
\usepackage{color}
\usepackage{graphicx}
\usepackage{natbib}

\def\apj{ApJ}%
\def\apjl{ApJ}%
\def\apjs{ApJS}%
\def\aap{A\&A}%
\def\mnras{MNRAS}%
\def\na{New A}%
\def\nat{Nature}%
\title[Stellar GADGET]{Stellar GADGET: A smooth particle hydrodynamics code for stellar astrophysics
and its application to Type Ia supernovae from white dwarf mergers}
\author[R.~Pakmor et al.]
       {R.~Pakmor$^{1,2}$, P.~Edelmann$^2$, F.~R\"{o}pke$^{3,2}$, W.~Hillebrandt$^2$ \\
        $^{1}$Heidelberger Institut f\"{u}r Theoretische Studien, Schloss-Wolfsbrunnenweg 35, D-69118 Heidelberg, Germany \\
        $^{2}$Max-Planck-Institut f\"{u}r Astrophysik, Karl-Schwarzschild-Str. 1, 85741 Garching, Germany \\
        $^{3}$Universit{\"a}t W{\"u}rzburg, Am Hubland, 97074 W{\"u}rzburg, Germany }

\date{Accepted.
      Received.}

\begin{document}
  \maketitle

  \label{firstpage}

  \begin{abstract}
    Mergers of two carbon-oxygen white dwarfs have long been suspected to be progenitors of Type Ia Supernovae.
    Here we present our modifications to the cosmological smoothed particle hydrodynamics code \textsc{Gadget} 
    to apply it to stellar physics including but not limited to mergers of white dwarfs. We demonstrate a new method
    to map a one-dimensional profile of an object in hydrostatic equilibrium to a stable particle distribution.
    We use the code to study the effect of initial conditions and resolution on the properties of the merger of
    two white dwarfs. We compare mergers with approximate and exact binary initial conditions and find that exact
    binary initial conditions lead to a much more stable binary system but there is no difference in the properties
    of the actual merger. In contrast, we find that resolution is a critical issue for simulations of white dwarf mergers.
    Carbon burning hotspots which may lead to a detonation in the so-called violent merger scenario emerge only
    in simulations with sufficient resolution but independent of the type of binary initial conditions. We conclude
    that simulations of white dwarf mergers which attempt to investigate their potential for Type Ia supernovae should
    be carried out with at least $10^6$ particles.
  \end{abstract}


  \section{Introduction}
    Mergers of two white dwarfs have first been proposed as progenitor systems of Type Ia Supernovae (SNe Ia) by
    \citet{iben1984a} and \citet{webbink1984a}. More than 25 years later, however, we are neither able to confirm them
    nor rule them out theoretically nor observationally.
    
    On the theory side there is a long tradition of numerical simulations of white dwarfs mergers to determine the fate of
    those systems after the merger. With time, these simulations have become better and more and more sophisticated
    both in terms of the numerical resolution used and the treatment of the input physics.
    
    The merger of two white dwarfs is an inherently three-dimensional problem. There is no intrinsic symmetry that
    can be exploited to simulate it in less then three dimensions.
    Starting with the pioneering work by \citet{benz1990a} most simulations of white dwarf mergers used smoothed
    particle hydrodynamics (SPH) codes. This results from several advantages SPH codes have compared to finite
    volume codes for this particular problem, including the fact that SPH conserves angular momentum very well, does not
    need a description of the volume surrounding an object and offers a simple way of implementing additional physics.
    Nevertheless, there are
    a few attempts to use finite volume codes \citep{dsouza2006a, motl2007a}. A problem of these simulations,
    however, has been shown to be the use of a polytropic equation of state which suppresses shocks and leads to an
    artificial increase of the orbital separation \citep{dan2011a}.
    Qualitatively, their results were not so different from \citet{benz1990a}, although they used a resolution of
    only $7000$ particles. They showed that the merger of two white dwarfs will lead to the destruction of the
    secondary, less massive, white dwarf. Its material forms a hot envelope and an accretion disk around
    the remaining primary white dwarf.
    
    The result of the long-term evolution of such a merger remnant is then likely to be the conversion of the central
    core into an oxygen-neon white dwarf by a slowly inward propagating flame \citep[e.g.][]{nomoto1985a,saio1998a}.
    If this oxygen-neon white dwarf accretes enough material from its envelope and the accretion disk to reach the
    Chandrasekhar mass, it will undergo an accretion-induced collapse, rather than a thermonuclear explosion
    \citep{nomoto1991a}.
    
    Following \citet{benz1990a}, merger simulations have been repeated with more and more particles (i.e.
    \citet{guerrero2004a} using $4 \times 10^4$ particles, \citet{yoon2007a} using $2 \times 10^5$ particles,
    and \citet{loren2009a} using $4 \times 10^5$ particles) confirming previous results. Starting with \citet{guerrero2004a}
    a nuclear reaction network has been coupled to the equations of hydrodynamics to account for the energy release from
    nuclear reactions during the merger.
    
    Only recently, indications have been found that those results may not be the full picture. \citet{pakmor2010a} who
    simulated the merger of two equal-mass white dwarfs with a mass ratio of one with an unprecedented resolution of
    $2 \times 10^6$ particles found that in the final phase of the merger, just before the disruption of the secondary
    white dwarf, a hotspot develops on the surface of the primary white dwarf. In this hotspot carbon burning starts
    and a detonation may form that directly consumes the whole merging binary system leading to a thermonuclear
    explosion.
     
    This violent merger scenario, of course, stands and falls with the assumption that a detonation forms. Further
    work showed that these hotspots form even with lower mass ratios of the merging white dwarfs, down to
    $0.8$ \citep{pakmor2011b}.
    Doubts on the formation of those hotspots have been raised by \citet{dan2011a} who introduced a method to obtain
    more relaxed (``exact'') binary initial conditions. They found that this leads to a significantly more stable binary system and
    without hotspots at a resolution of $2 \times 10^5$ particles. In contrast, \citet{raskin2012a} and \citet{pakmor2012a} found
    hotspots to be present in merger simulations with exact binary initial conditions and a resolution of $10^6$ particles
    and more. Note, however, that all these simulations have been run with different codes, different parameters, and
    at least partially different input physics.
     
    In this paper, we therefore describe in detail our methodology, which is based on the \textsc{Gadget} code. While we emphasize
    that the modifications of the code, that originally was designed to address problems in cosmology, are useful for a potentially
    wide range of problems in stellar astrophysics, we use our implementation to test the role of initial conditions and resolution in
    mergers of white dwarfs here.
    
    In Sec.~\ref{sec:code} we describe in detail the modification we made to the \textsc{Gadget} code and test our new timestepping mechanism
    in Sec.~\ref{sec:wakeup}. We present our new nuclear reaction network in Sec.~\ref{sec:network}. We discuss our new method to 
    create initial conditions for stars in SPH in Sec.~\ref{sec:icsall} and demonstrate that it works well.
    In addition, we describe our implementation to create ``exact'' binary initial conditions following \citet{dan2011a}.
    We then show results of simulations of merging white dwarfs in Sec.~\ref{sec:mergers}, studying in particular
    the influence of binary initial conditions and resolution on the merger of two massive white dwarfs using the same
    code and parameters for all simulations. We finish with a summary of the paper and some conclusions in Sec.~\ref{sec:summary}.

  \section{Code modifications}
    \label{sec:code}
    A detailed description of the \textsc{Gadget} code can be found in \citet{springel2005a} and \citet{springel2001a}.
    Our changes include additional physics as well as technical changes to adapt the code to
    different conditions. The changes are as follows.

    \begin{enumerate}
    \item \textit{Gravitational softening}\\
    To calculate the gravitational forces we use the standard tree algorithm implemented in \textsc{Gadget}.
    Instead of a fixed gravitational softening length as usually applied in cosmological simulations, we use the individual
    smoothing lengths of gas particles as their gravitational softening \citep{bate1997a}. This choice significantly improves the
    stability of objects in hydrostatic equilibrium. However, it can lead to errors in the total energy budget of the
    simulation when the smoothing length of a particle changes, as the gravitational softening and therefore the
    local gravitational potential also changes. As described by \citet{price2007a}, it is possible to compensate for
    this by adding extra terms to the evaluation of the gravitational force (for an implementation into the 
    \textsc{Gadget} code see \citet{iannuzzi2011a}). The improved energy conservation, however, comes at the cost of
    additional noise in the velocity field. Therefore, we refrain from using it.
    For most applications the errors introduced are small anyway, because the smoothing lengths of the particles
    do not change significantly during the simulation. \\

    \item \textit{Wakeup mechanism}\\
    In the \textsc{Gadget} code, individual time steps are assigned to all particles, which depend only on the local
    conditions near the particles. The particles are then evolved on their own time steps. While this greatly reduces the computational
    costs to run a simulation, it can cause severe problems if very fast particles run into a medium in which the sound speed is much
    smaller then the velocity of the fast particles. In this case, a fast particle
    on a small time step can pass through a particle on a much larger time step without being noticed. As this
    behavior can obviously lead to completely unphysical results, we implement a ``wakeup mechanism'' for the time-stepping.
    Similar to the method proposed by \citet{saitoh2009a}, it should activate inactive particles as other particles approach
    them which evolve on much shorter time steps. \\
    The hydrodynamical time step of an SPH particle is calculated as
    \begin{equation}
      \Delta t_i = \frac{C_{\mathrm{courant}}\ h_i}{\max_j( v^{\mathrm{signal}}_{ij} )}
    \end{equation}
    where $C_{\mathrm{courant}}$ is the Courant factor, $h_i$ is the smoothing length of the particle and the denominator 
    finds the maximum of the signal velocities $v^{\mathrm{signal}}_{ij}$ from particle $i$ and all its neighbors $j$ as
    defined in equation~13 of \citet{springel2005a}. The resulting maximum signal velocity of a particle is stored. In each
    time step we check for all inactive neighbors of all active particles whether their stored maximum signal velocity is
    smaller by some factor than the new signal velocity calculated for the active particle and the inactive neighbor,
    \begin{equation}
      v^{\mathrm{maxsignal}}_{i} < w \cdot v^{\mathrm{signal}}_{ij}.
    \end{equation}
    If fulfilled, we flag the particle to be woken up. Our usual choice for the wakeup factor $w$ is $4.1$; thus an active particle
    can be in range of an inactive particle for three time steps at most. The check can be done
    very efficiently, since the active particle has to loop over all its neighbors anyway to calculate the local pressure
    force. After finishing the time step, we change the time step of all particles that have been flagged to wake up such that
    they become active in the next possible time step. When a particle was active in a time step, its properties had been predicted
    half a time step into the future using its actual rates of change of these properties. Thus, this extrapolation needs to be corrected
    when we shorten the time step. The correction for the difference between the estimated time step and the time step the particle really
    experienced is done consistently for all quantities. \\

    \item \textit{Energy equation}\\
    In contrast to the original implementation of \textsc{Gadget} that uses the entropy equation we solve the energy equation. 
    This is convenient choice because evolving the internal energy simplifies combining hydrodynamics with the nuclear
    reaction network. \\

    \item \textit{Equation of state}\\
    The equation of state (EoS) captures all intrinsic properties of the material. It is used to calculate local pressure and
    speed of sound from the primary thermodynamical variables evolved in the code. The original \textsc{Gadget} code only
    implements an ideal gas equation of state. We replace it with the Helmholtz-EoS \citep{timmes2000a}. This EoS describes an
    arbitrarily degenerate, arbitrarily relativistic electron--positron gas together with an ideal gas of completely ionized ions. It also includes radiation from a
    black body with the local gas temperature.
    Since internal energy is our thermodynamic variable of choice, most calls of the EoS have the
    internal energy as input. Because it is in general not possible to invert the EoS in a closed form, we need to iterate on the
    temperature first, using a Newton method. 
    Afterwards all other quantities are calculated from temperature, density and composition. Due to numerical errors (e.g.\ when the value of the
    internal energy becomes smaller than the minimum energy of the degenerate electron), it may not be possible to find a valid
    temperature for a given internal energy. In this rare case, or when the temperature drops below the minimum temperature tabulated by
    the EoS, we assign a temperature of $1000 \mathrm{K}$ to the particle. All other thermodynamical quantities apart from the internal
    energy are then calculated for this temperature. \\
   
    \item \textit{Nuclear reaction network}\\
    Because SPH is a purely Lagrangian method, the composition of a particle can only change by nuclear reactions. The nuclear network
    is coupled to hydrodynamics in an operator splitting approach. After a hydrodynamics timestep, the nuclear network is integrated
    for all active particles for the duration of their timestep. During this integration, the density is assumed to be constant.
    Changes of the temperature as a result of energy release or consumption by nuclear reactions are taken into account in the network.
    The nuclear reaction network and its integration are described in detail in Section~\ref{sec:network}.
    We restrict evolving the nuclear network to particles with temperatures higher than $10^7\, \mathrm{K}$. For most of our applications
    this is a very conservative estimate, as carbon burning starts only at around $10^9\, \mathrm{K}$.

    After evolving the nuclear network for the duration of the hydrodynamical time step, the composition of the particle is changed according
    to the nuclear reaction network. From the rate of change of the abundances of the different species we then calculate the amount of energy
    that is released or consumed by nuclear reactions.
    \begin{equation}
       \dot{e}_\mathrm{nuc} = \sum_{j=1}^{N_\mathrm{species}} N_{A}\ M_j c^2\ \frac{\mathrm{d}Y_j}{\mathrm{d}t}
       \label{eqn:energy-release}
    \end{equation}
    Here, $N_{A}$ is Avogadro's constant, $M_j c^2$ is the rest mass energy of species~$j$, and $Y_j$ is its number fraction. This change of energy is included as a source
    term in the energy equation. \\
    \end{enumerate}

     \begin{figure}
      \centering
      \includegraphics[width=0.9\linewidth]{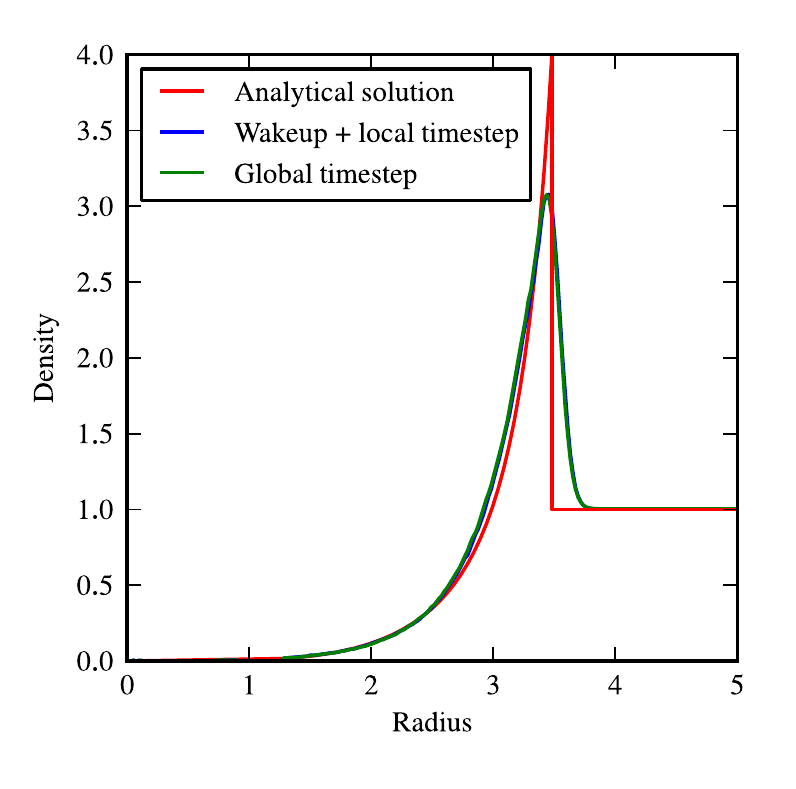}
      \caption[Sedov problem]
      {Radial density profiles of two simulation runs of the Sedov problem with global timesteps and local timesteps with the wakeup mechanism, respectively. 
       The analytical solution is shown for comparison.}
      \label{fig:sedov}
    \end{figure}   
 
  \section{Wakeup test: Sedov Problem with local timestepping}
    \label{sec:wakeup}
    To check the accuracy of our wakeup mechanism we apply it to the Sedov problem \citep{sedov1959a}, which is a simple point explosion leading to a strong
    shock wave expanding in a surrounding medium. We start from glass initial conditions \citep{white1994a} in a periodic box of size $10$ in arbitrary units that 
    contains $10^6$ particles of with a mass of $10^{-3}$ each.
    Then we inject an internal energy of $10^5$ equally into the $33$ particles closest to the center of the box. The internal energy
    of the surrounding particles is chosen to be $10^{-5}$. For this configuration there is a large spread of local timesteps. Particles which are influenced by one
    of the hot particles (i.e.\ one of them is closer than their smoothing length) are evolved on a timestep of $10^{-5}$. All other particles, however, start with a timestep of
    $5 \times 10^2$, more than a million times larger. Therefore, these particles obviously decouple completely from the hot particles and their close neighbours.
    Although the conditions we use are quite extreme, the same problem also occurs for real problems, i.e.\ the interaction of the ejecta of a supernova explosion
    with a companion star in a binary system \citep[e.g.][]{pakmor2008a}. One way to circumvent this problem is to use the same global timestep for all particles, which
    guarantees correct behavior. However, it makes the simulation significantly more expensive, because we then have to evolve all particles on a very small timestep
    from the beginning rather than from the time when the shock actually hits them. Another approach is discussed in Sec.~\ref{sec:code}, which reduces the timestep
    of a particle as soon as the shock touches it. Fig.~\ref{fig:sedov} compares the radial density profile for the Sedov problem for a global timestep and a local
    timestep combined with this wakeup mechanism. It clearly shows that using local timesteps plus the wakeup mechanism recovers exactly the same result
    we get from using a global timestep.

  \section{The nuclear network}
    \label{sec:network}
    The nuclear reaction network calculates the change of the composition with time. The change of the abundance $\mathrm{Y}_i$ of one species depends on the
    reaction rates $\mathcal{R}_k$ that destroy or create nuclei of this species. 
    \begin{equation}
       \dot{\mathrm{\textbf{Y}}}_i = \sum_k \mathcal{R}_k \left( \rho, T, \textbf{Y} \right )
       \label{eqn:rates}
    \end{equation}
    Each reaction rate depends on the local density, temperature and abundance of the reactants. The reaction rates are tabulated and
    taken from the latest~(2009) release of the REACLIB database \citep{rauscher2000a}, which includes experimental data as well as theoretical rates
    when experimental data are not available. Additionally, the weak interaction rates of \citet{langanke2001a} are included. Neutrino losses are neglected.

    Under the conditions relevant for nucleosynthesis atoms are usually fully ionized.
    Nevertheless, the assumption of bare nuclei colliding with each other is not valid, because the electron background shields part of the charge of
    the nuclei. Thus, since the Coulomb repulsion is reduced, the nuclei can interact more easily and the reaction rate increases. We
    treat the effects of electron screening on the reaction rates as described by \citet{wallace1982a}. This description discriminates between regimes of strong \citep{alastuey1978a} and weak \citep{dewitt1973a} screening
    and an intermediate region. In contrast to the original paper, we use a different interpolation 
    method in the intermediate regime and a corrected coefficient for strong screening, both taken from \citet{wallace1983a}.

    Since the reaction rates depend on temperature to a high power and temperature can change
    quickly due to the energy release of nuclear reactions, the temperature has to be evolved along with the abundances for correct integration.
    This is taken care of by an additional equation for the evolution of temperature
    \begin{equation}
      \dot{T} = \left. \dot{E} \frac{\partial T}{\partial E} \right \vert_{\rho=\mathrm{const}} + \sum_i \dot{Y_i} \frac{\partial T}{\partial Y_i}.
    \end{equation}
    It should be noted that $T$ is completely determined by the abundances~$Y_i$ and the density, if the energy release is given by
    equation~(\ref{eqn:energy-release}). Therefore it is not necessary to include it explicitly as an additional variable. The problem with this
    approach is that the Jacobian of the right side of equation~(\ref{eqn:rates}) would be a dense matrix. By separating the temperature evolution into an
    additional equation, the Jacobian becomes sparse, with non-zero entries at the respective input and output nuclei of the corresponding reactions.
    Therefore, we choose to include the extra equation for $T$ to be able to use significantly faster algorithms for inverting the Jacobian.
    
    We assume the density to be constant over one timestep. Once this approximation is not valid anymore, an evolution equation
    for the density can be implemented the same way as for the temperature. Some kind of thermodynamic constraint (e.g.\ constant pressure) is needed in this case.

    If electron screening is accounted for, the same method is used for the quantity~$\sum_{j=1}^{N_\mathrm{species}}{Z_j}^2 Y_j$ on which the
    screening factors depend. The same would apply to $Y_\mathrm{e} = \sum_{j=1}^{N_\mathrm{species}}{Z_j} Y_j$ but because $Y_\mathrm{e}$ only
    changes slowly over a typical timestep it can safely be assumed be constant for the computation of the Jacobian.
 
    Thus, we end up with a system of $N + 1$ ($N+2$, with screening) ordinary differential equations for $N + 1$ ($N+2$, with screening) variables in a nuclear network of $N$~species. Because this system of equations is very stiff, it needs to be integrated implicitly. To this end we apply the variable-order
    Bader--Deuflhard method \citep{bader1983a} as suggested by \citet{timmes1999b}. Depending on the number of nuclei we use a full direct matrix
    solver (for small networks) or the sparse matrix solver \textsc{SuperLU} \citep{demmel1999a}.

    \begin{figure*}
      \centering
      \includegraphics[width=\textwidth]{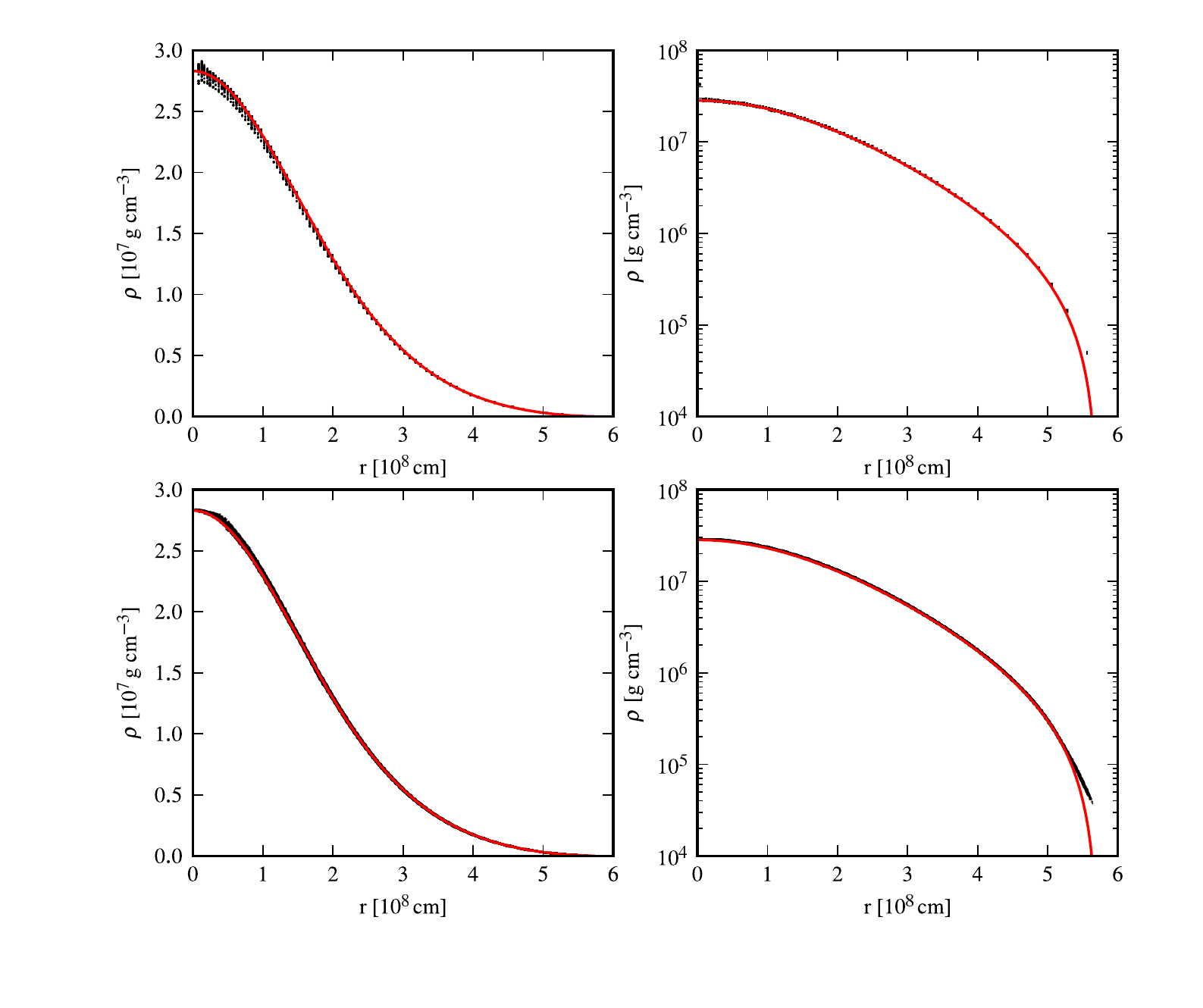}
      \caption[Initial conditions before and after relaxation]
      {Comparison of the initial conditions of a $1.0\,\mathrm{M_\odot}$ white dwarf generated by the \textsc{healpix}
      method (see text). Black points show the SPH density estimate of all particles depending on their
      radial coordinate. The red line shows the initial one-dimensional density profile. Left and right column show the
      same data, but use a linear/logarithmic scale for the density. The top row shows the initial setup, the
      bottom row shows the configuration after completion of the relaxation.}
      \label{fig:ic}
    \end{figure*}

  \section{Initial conditions for compact objects}
    \label{sec:icsall}
    Models of stars and compact objects usually assume hydrostatic equilibrium and spherical symmetry. When we want to simulate such objects
    in \textsc{Gadget}, we therefore have to start from one-dimensional profiles of density, composition, pressure, etc.\ in hydrostatic equilibrium.
    We then have to find a way to construct a stable three-dimensional particle distribution, that reproduces these profiles.
    Because the local density is coupled to the local particle distribution, noise in the density estimate of any non-trivial initial particle configuration
    cannot be avoided. It is essential to reduce this noise to maintain the initial configuration as accurate as possible.
    
    \subsection{Setup}
    \label{sec:ics}
    The idea of our method to construct the particle distribution of a single star is to divide the star into spherical shells and these shells into
    subvolumes of roughly cubical size to which we attribute one SPH particle each. In addition, each of the subvolumes should contain the same
    mass, to fulfill the constraint of equally massive particles. With this method we obtain a rather regular particle distribution that resembles
    any given radial density profile with only very small noise compared to a random sampling of the density profile.

    In our implementation we use the \textsc{healpix} library \citep{gorski2005a} to tessellate the surface of a sphere into $12 \cdot N^2$ 
    approximately quadratical pieces of the same area where the index $N$ is a non-negative integer. We then construct the star
    out of several shells starting from the center. The width of the shells is chosen such that it is of about the same size as the edges of
    the pieces. As we show below, this condition can be satisfied using a simple method to determine the width of the shells.
    We express the constraints using parameters $n_1$ and $n_2$, which are equal if both constraints are fulfilled. Then
    $N = round(n_1) = round(n_2)$ is used to tesselate the shell.

    Let us assume we know the lower radius $r_{\mathrm{lower}}$ of a shell and want to find its upper radius $r_{\mathrm{upper}}$. For a
    given upper radius, the width of a shell and its mass are given by
    \begin{equation}
      d_{\mathrm{shell}} = r_{\mathrm{upper}} - r_{\mathrm{lower}}
    \end{equation}
    and
    \begin{equation}
      m_{\mathrm{shell}} = 4\pi \int_{r_{\mathrm{lower}}}^{r_{\mathrm{upper}}} \rho(r) r^2 \mathrm{d}r,
    \end{equation}
    respectively. The mass increases with increasing $r_{\mathrm{upper}}$. Fixing the uniform particle mass in the beginning, we can calculate the number
    of particles that need to be placed in this shell to
    \begin{equation}
      n_{\mathrm{particles}} = \frac{ m_{\mathrm{shell}} }{ m_{\mathrm{particle}} }.
    \end{equation}

    This is equivalent to an index $n_1$ of
    \begin{equation}
      n_1 = \sqrt{ \frac{ m_{\mathrm{shell}} }{ 12 \cdot m_{\mathrm{particle}} } },
    \end{equation}
    which increases with increasing width of the shell.

    The second constraint on our shell is the requirement of cubical volumes. The surface area of a shell is roughly given by
    \begin{equation}
      S = 4 \pi r^2 = 4 \pi \left[ 0.5 \cdot \left( r_{\mathrm{lower}} + r_{\mathrm{upper}} \right) \right]^2.
    \end{equation}
    As each shell contains $12 \cdot n_2^2$ particles, the edge size of one piece on the surface can be written as
    \begin{equation}
      a_{\mathrm{particle}} = \sqrt{ \frac{S}{12 \cdot n_2^2} } = \sqrt{\frac{\pi}{12}} \left( r_{\mathrm{lower}} + r_{\mathrm{upper}} \right) \frac{1}{n_2}.
    \end{equation}
    This edge size should be equal to the width of the shell. Setting it equal to $d_\mathrm{shell}$ and solving for $n_2$ leads to
    \begin{equation}
      n_2 = \sqrt{\frac{\pi}{12}} \frac{ r_{\mathrm{lower}} + r_{\mathrm{upper}} }{ r_{\mathrm{upper}} - r_{\mathrm{lower}} }.
    \end{equation}
    While increasing the upper radius of the shell, the second index decreases. Since $n_1 = 0$ and $n_2 \rightarrow \infty$ for $r_{\mathrm{upper}} = r_{\mathrm{lower}}$
    it is always possible to increase the upper radius until $n_1$ equals $n_2$. Having found the upper radius, we place a shell of particles at 
    $r = 0.5 \cdot \left( r_{\mathrm{upper}} + r_{\mathrm{lower}} \right)$ using the coordinates from the \textsc{healpix} library and continue with the next shell. 
    Internal energy and composition of the particle are chosen according to the radial coordinate of the particles. The initial velocity of all particles is
    set to zero.

  \subsection{Relaxation}
    \label{sec:relax}
    To damp out spurious numerical noise introduced by the setup, we relax the object before we start the actual simulation. To this end we evolve it for several dynamical
    timescales while applying a time-dependent damping force similar to that used by \citet{rosswog2004a}. The specific force is given by
    \begin{equation}
      \dot{\textbf{v}}_i = - \frac{ \textbf{v}_i }{ \tau }
    \end{equation}
    and applied together with gravitational and hydrodynamical forces. The damping timescale $\tau$ controls the strength of the damping term. The smaller it is,
    the stronger is the damping. We start with a large damping force that decreases with time and is eventually switched off completely. Afterwards we continue the simulation
    for a short time. If the relaxation has been successful, the object will remain in equilibrium and particle motions will stay close to zero. Otherwise, assuming
    that the initial model is not intrinsically unstable, the relaxation parameters have to be adjusted. We vary the damping timescale with time as
    \begin{equation}
      \frac{1}{\tau} = \left\{
      \begin{array}{c@{\quad}l}
      \displaystyle \frac{1}{\tau_0} & \displaystyle t \le 0.2\ t_{\mathrm{max}} \\
      & \\
      \displaystyle \frac{1}{\tau_0 \cdot 10^\frac{3(t - 0.2\ t_{\mathrm{max}})}{ 0.6\ t_{\mathrm{max}} } } & \displaystyle 0.2\ t_{\mathrm{max}} \le t \le 0.8\ t_{\mathrm{max}} \\
      & \\
      \displaystyle 0 & \displaystyle t > 0.8\ t_{\mathrm{max}}
      \end{array}
      \right.
    \end{equation}
    We run the relaxation for a total time $t_{\mathrm{max}}$, which is chosen to be at least several dynamical timescales. The initial damping timescale $\tau_0$
    should be much smaller than the dynamical timescale. Typical choices for a white dwarf mass of $1.0 \mathrm{M_\odot}$ are $\tau_0 = 0.002 \mathrm{s}$ and
    $t_{\mathrm{max}} = 100 \mathrm{s}$. Figure~\ref{fig:ic} shows the configuration of the initial setup and after the relaxation step. The final relaxed configuration
    agrees nearly perfectly with the initial profile. Only for the very outermost particles of the star the density is overestimated systematically, because these
    particles only have neighbors which are at smaller radii and have a higher density.
    
  \subsection{Binary initial conditions}
    \label{sec:binics}
    Having obtained equilibrium models for individual stars we need to join two of them to form a binary system. For this we use two different approaches, which
    we will label \textit{approximate} and \textit{exact} binary initial conditions, following \citet{dan2011a}. For \textit{approximate} initial conditions both stars
    are set on a circular orbit with an orbital period to render the binary system marginally unstable. The initial orbital period is not determined exactly and can be chosen
    in different ways. Those include an iterative procedure until the system is stable for a certain number of orbits or using an approximation for the distance at the onset
    of Roche lobe mass transfer \citep{eggleton1983a}.
    To construct \textit{exact} binary initial conditions we implement the approach used by \citet{dan2011a}. We first set both stars on a stable, 
    circular, synchronized orbit with a large orbital period such that there is no mass transfer and the binary system is stable. Then we evolve the binary system in the
    co-rotating frame and add an artificial damping force to remove all residual velocities. This is done by adding an additional acceleration $\textbf{a}_{\mathrm{ext},i}$ on each particle
    that is given by
    \begin{equation}
      \textbf{a}_{\mathrm{ext},i} = -\, \mathbf{ \omega } \times \left( \mathbf{\omega} \times \mathbf{r}_i \right) - 2\, \mathbf{\omega} \times \mathbf{v}_i - \frac{\mathbf{v}_i}{\tau_\mathrm{damp}},
    \end{equation}
    where $\mathbf{r}_i$ and $\mathbf{v}_i$ are position and velocity of particle $i$ in the co-rotating frame, $\mathbf{\omega}$ the orbital frequency of the binary system,
    and $\tau_\mathrm{damp}$ the damping timescale. The orbital frequency is updated in each global timestep, i.e.\ when all particles are active at the same time. It is calculated
    as the average of the orbital frequencies necessary to exactly balance the total force on the two individual stars. The difference between the two individual orbital
    frequencies calculated for both stars is typically smaller than one percent. For the white dwarf merger simulations presented in this work we set $\tau_\mathrm{damp}$ to
    $0.005\,\mathrm{s}$, about a factor of five larger than the smallest timesteps in the high resolution run with $1.6 \times 10^6$ particles.
    
    We then apply an artificial radial drift velocity to both stars such that the center of mass of the binary system does not move. The drift velocity is estimated as described in
    Equation (8) of \citet{dan2011a}. We continue the simulation until the first particle of the less massive white dwarf crosses the inner Lagrange point. At this point we switch
    off damping and radial drift, and use the current orbital frequency to transform the binary system into a non-rotating frame. The resulting binary system then gives us the
    initial conditions for simulations of merging white dwarfs as presented in Section~\ref{sec:mergers}.
    
   \begin{table*}
      \caption{Model parameters and properties of the four merger simulations. Tabulated are the total number of particles in the simulation $\mathrm{N_{p}}$, the
      type of binary initial conditions used (IC Type), the orbital period $T$, orbital separation $a$ and total angular momentum $L$ as properties of the initial binary
      system. In addition, several properties of the binary system are shown at the time $t$ when the separation is $6 \times 10^8\,\mathrm{cm}$. These are the number
      of particles $\mathrm{N_{hot}}$ on the surface of the primary with a temperature larger than $2 \times 10^9\,\mathrm{K}$, the central densities of the primary
      and the secondary  white dwarf ($\mathrm{\rho_{sec,max}}$ and $\mathrm{\rho_{prim,max}}$, respectively) and an estimate of the smoothing length in the hotspot
      on the surface of the primary, if present.}
      \label{table:models}
      \begin{tabular}{lrccccccccc}
        \hline
        Number & $\mathrm{N_{p}}$ & IC Type & $T_\mathrm{initial}$ & $a_\mathrm{initial}$ & $L_\mathrm{initial}$ & $t$ & $\mathrm{N_{hot}}$ & 
        $\mathrm{\rho_{prim,max}}$ & $\mathrm{\rho_{sec,max}}$ & $\mathrm{Hsml_{hotspot}}$ \\
         & $\mathrm{[10^5]}$ & & $\mathrm{[s]}$ & $\mathrm{[10^9\,cm]}$ & $\mathrm{[10^{50}\,gcm^2s^{-1}]}$ & $\mathrm{[s]}$ & & $\mathrm{[10^7\,gcm^{-3}]}$ & 
         $\mathrm{[10^6\,gcm^{-3}]}$ & $\mathrm{[10^7\,cm]}$ \\
         \hline
        1 & $1.8$ & approx. & 28.0 & 1.74 & 7.18 & $73$   & $8$   & $5.6$ & $7.0$ & $5$ \\
        2 & $18$  & approx. & 28.0 & 1.74 & 7.18 & $77$   & $81$ & $6.1$ & $4.2$ & $3$ \\
        3 & $1.8$ & exact     & 32.6 & 1.93 & 7.57 & $469$ & $--$   & $5.2$ & $7.1$ & $--$ \\
        4 & $18$  & exact     & 32.6 & 1.93 & 7.55 & $615$ & $44$ & $5.5$ & $4.2$ & $3$ \\
        \hline
      \end{tabular}
    \end{table*}
    
    \begin{figure*}
      \centering
      \includegraphics[width=0.9\linewidth]{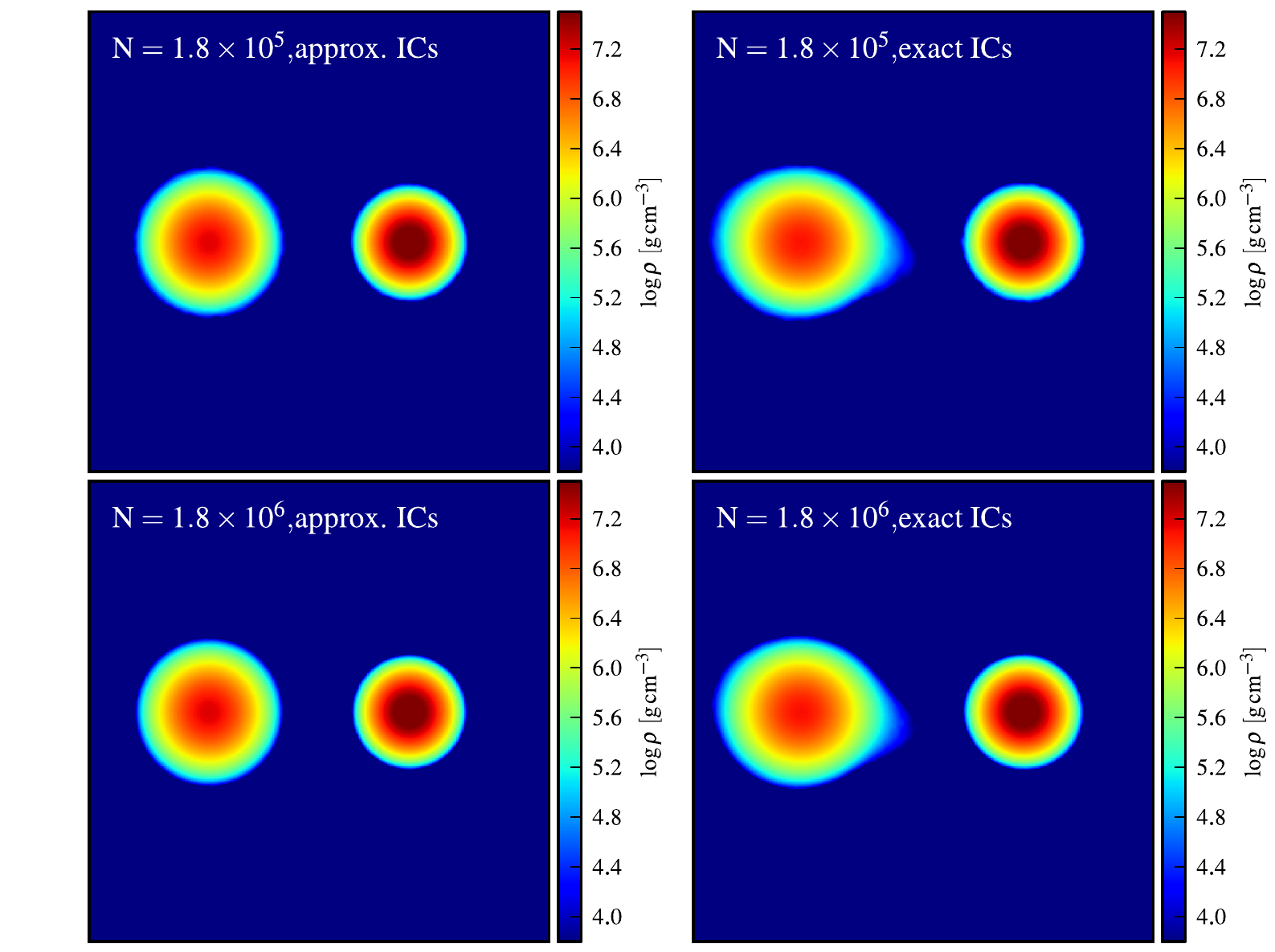}
      \caption{Initial conditions for the merger simulations. Shown is a density slice in the plane of rotation with a size of $4 \times 10^9\,\mathrm{cm}$.}
      \label{fig:binics}
    \end{figure*}
    
    \begin{figure}
      \centering
      \includegraphics[width=0.9\linewidth]{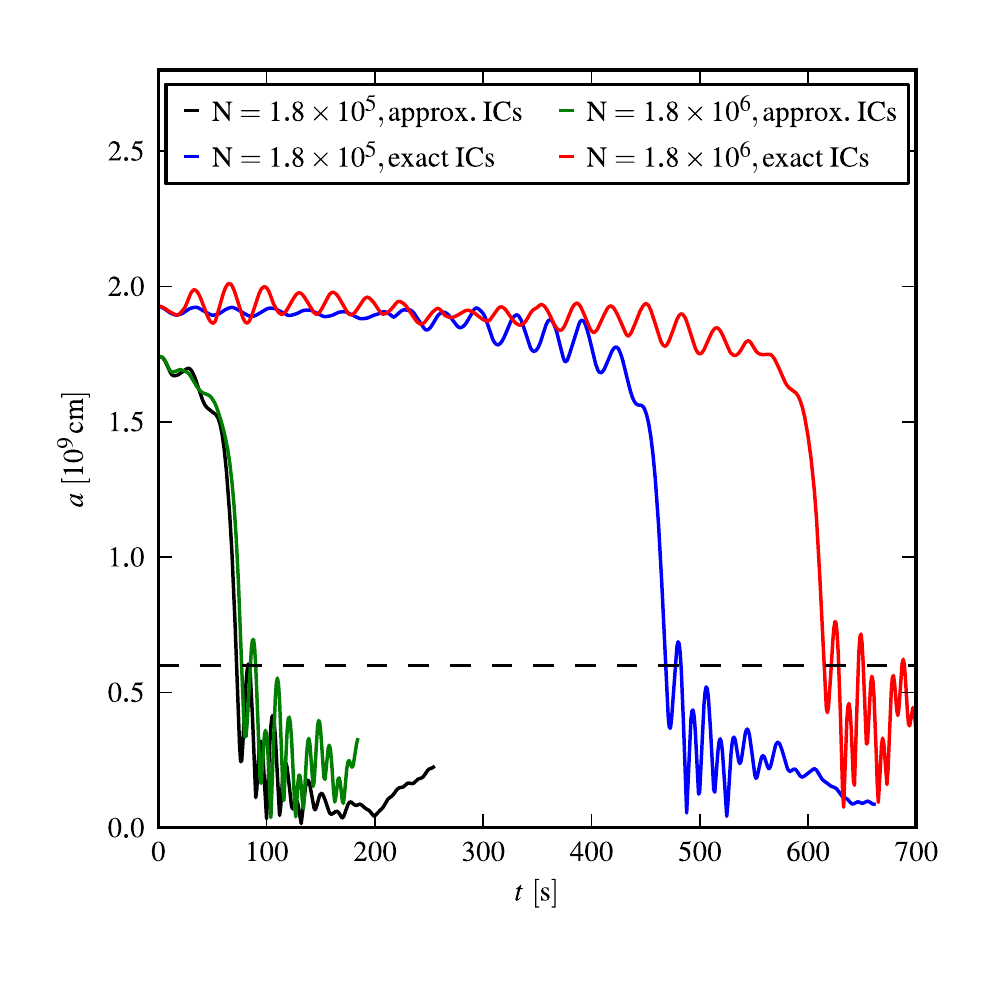}
      \caption{Evolution of the orbital separation for the merger models until the binary system has merged. The dashed black line shows the separation at which
      we compare the different simulations in detail.}
      \label{fig:dist}
    \end{figure}
    
  \section{Application to double white dwarfs mergers}
    \label{sec:mergers}
    \subsection{Setup}
    To study the influence of resolution and initial conditions on the properties of the merger simulation we pick a specific binary system of a $1.1\,\mathrm{M_\odot}$
    primary and a $0.9\,\mathrm{M_\odot}$ secondary carbon-oxygen white dwarf. This binary system is of particular interest, because it has been shown to lead to
    observables that closely resemble normal SNe Ia under the assumption that a detonation forms during the merger \citep{pakmor2012a}.
    
    Here we perform four merger simulations of this binary system with low ($1.8 \times 10^5$ SPH particles) and high ($1.8 \times 10^6$ SPH particles) resolution 
    and approximate and exact binary initial conditions for each, respectively. The properties of these models are summarized in Table~\ref{table:models}.
    The individual white dwarfs in these models are first constructed as one-dimensional models in hydrostatic equilibrium with a uniform temperature of $10^5\,\mathrm{K}$
    and composition of equal parts by mass of $^{12}\mathrm{C}$ and $^{16}\mathrm{C}$. We then create three-dimensional particle distributions resembling the
    one-dimensional profiles as described in Sec.~\ref{sec:ics}. The mass of the particles is chosen such that all particles of a particular binary system have the same mass.
    We relax the individual white dwarfs for $10\,\mathrm{s}$ as discussed in Sec.~\ref{sec:relax}.
    Two white dwarfs are then combined to a binary system as described in Sec.~\ref{sec:binics}. To obtain exact initial conditions we use a relaxation timescale of
    $5\times10^{-3}\,\mathrm{s}$ and a radial velocity of $1.0 \times 10^7 \,\mathrm{cm\,s^{-1}}$ for the $0.9\,\mathrm{M_\odot}$ white dwarf and
    $0.82 \times 10^7 \,\mathrm{cm\,s^{-1}}$ for the $1.1\,\mathrm{M_\odot}$ white dwarf. We start with an initial orbital period of $50\,\mathrm{s}$ and stop when the first
    particle of the secondary white dwarf crosses the inner Lagrange-point of the binary system. The orbital separation and the orbital period at this time are given in
    Table~\ref{table:models}. We then remove the radial and add the orbital velocity and restart the simulation in a fixed frame.
    
    \begin{figure*}
      \centering
      \includegraphics[width=0.9\linewidth]{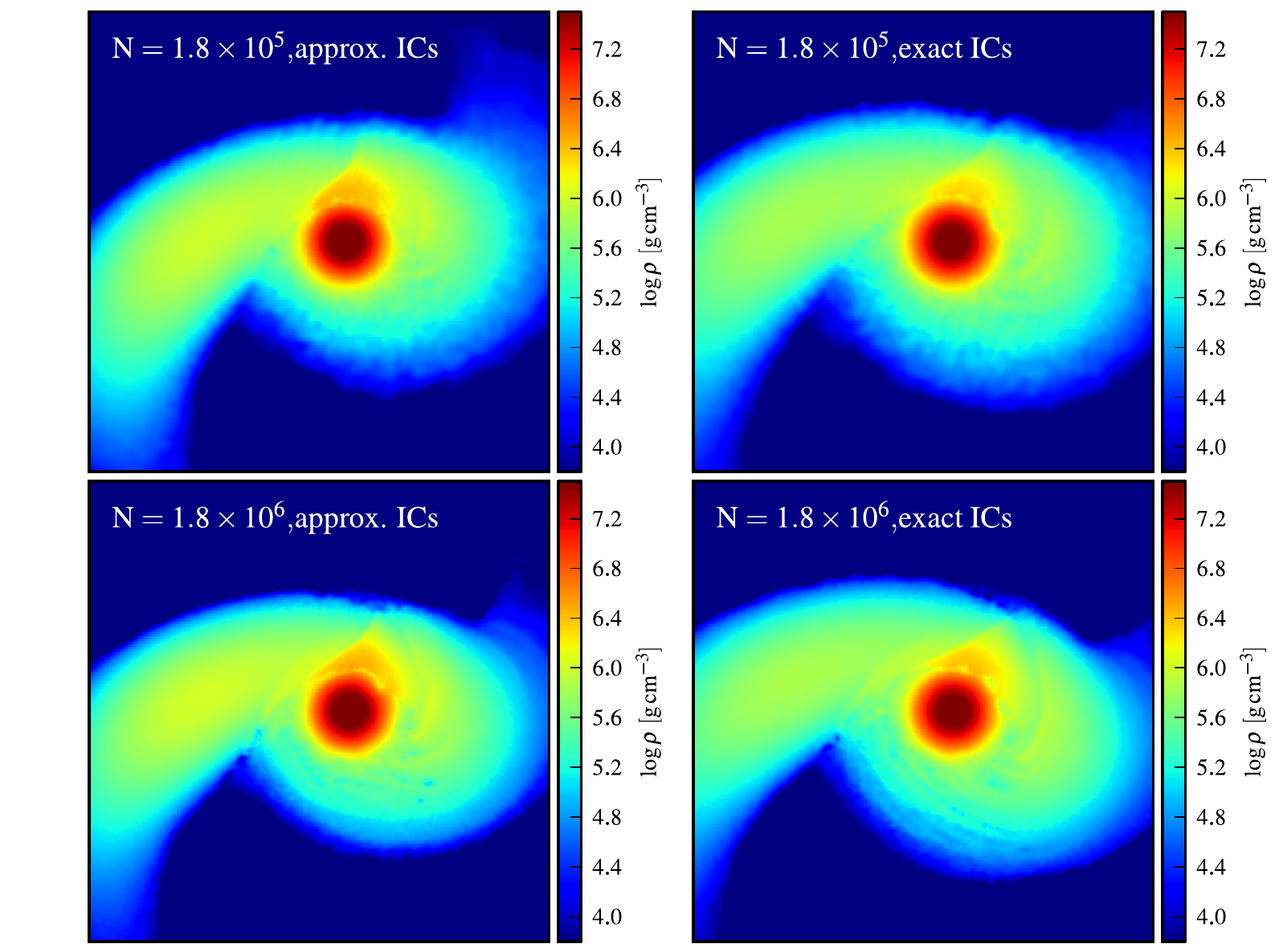}
      \caption{Density slice through the binary system when the separation distance is $6\times 10^8\,\mathrm{cm}$. The individual plots are centered on the center of
      mass of the binary system and rotated to have the centers of mass of both white dwarfs on the x-axis. The box has a height and width of $4 \times 10^9\,\mathrm{cm}$.}
      \label{fig:rho}
    \end{figure*}
    
    \begin{figure*}
      \centering
      \includegraphics[width=0.9\linewidth]{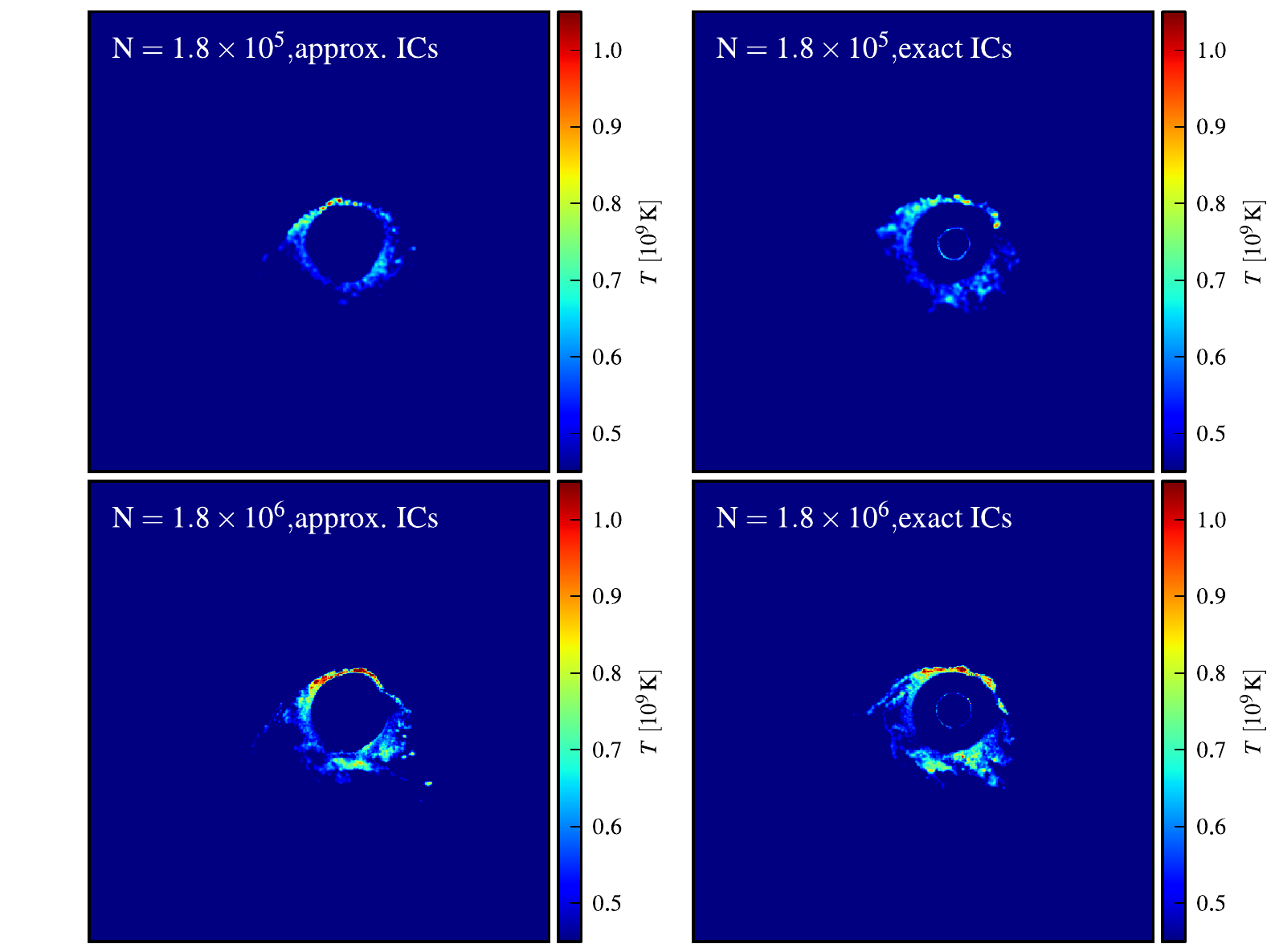}
      \caption{Temperature slice through the binary system when the separation distance is $6\times 10^8\,\mathrm{cm}$. The individual plots are centered on the center of
      mass of the binary system and rotated to have the centers of mass of both white dwarfs on the x-axis. The box has a height and width of $4 \times 10^9\,\mathrm{cm}$.}
      \label{fig:temp}
    \end{figure*}
        
    The density distribution of the initial conditions is shown in Fig.~\ref{fig:binics}. In the models with exact initial conditions, the secondary white dwarf is slightly deformed
    and fills its Roche-lobe. The initial separation of the models with exact initial conditions is a bit larger compared to the models with approximate initial conditions, which
    also gives them a slightly larger (by $\approx 5\%$) total angular momentum. The resolution has no significant effect on the initial conditions.
    
    The configuration of the \textsc{Gadget} code is mostly identical for all four simulations. We use the nuclear reaction network described in Sec.~\ref{sec:network} for
    the 13 $\mathrm{\alpha}$-element isotopes from $^{4}\mathrm{He}$ to $^{56}\mathrm{Ni}$. It is active only for particles with a temperature
    larger than $5\times10^8\,\mathrm{K}$, because carbon burning is negligible at lower temperatures. In the high resolution simulations we limit the smoothing
    lengths of the particles to be smaller or equal to $2\times10^8\,\,\mathrm{cm}$. This only affects less than $1\%$ of the particles, in particular the particles which are
    ejected from the binary system during the inspiral phase.
  
  \subsection{Inspiral}
    The main difference between the simulations in the inspiral phase is in the stability of the binary systems and thus the time it takes until the binary system merges.
    Fig.~\ref{fig:dist} shows the evolution of the orbital separation for all four simulations. Obviously, there is a much larger difference between models with different
    binary initial conditions than between models with different resolution. In particular, binary systems with exact initial conditions are significantly more stable than
    binary systems with approximate initial conditions, in agreement with previous results by \citet{dan2011a}. In our case, the binary systems with approximate
    initial conditions are only stable for $2$ to $3$ orbits before they merge. Their stability shows basically no dependence on resolution. In contrast, the binary systems
    with exact initial conditions are stable for about $15$~orbits (the lower resolution simulation) and nearly $20$~orbits (the high resolution simulation).
    It is important to note that although the stability of the binary systems is quite different the final merger occurs on the timescale of about one orbit in all simulations.
    Note also that there are systematic problems with SPH, which can affect the mass accretion rate and the stability of the binary system. As mentioned before, 
    the density of the very outermost particles of the two objects is overestimated systematically. In addition, there is a considerable artificial surface tension. Both effects
    tend to reduce mass transfer and therefore can stabilize the binary system. In the end, only high-resolution simulations with different hydrodynamical schemes will
    be able to show how long the mass transfer phase before the actual merger really takes.
 
  \subsection{The last binary orbit}
    The most important stage for the fate of the binary system is the final orbit of the merger when the secondary white dwarf is destroyed. At this time, it is decided
    whether a detonation forms which leads to the violent merger scenario \citep{pakmor2010a,pakmor2011b,pakmor2012a} or whether the system ultimately forms a cold
    primary white dwarf surrounded by an accretion disk and a hot envelope made from the material of the disrupted secondary white dwarf \citep[see, e.g.][]{dan2011a}.
    
    To investigate the effect on resolution and binary initial conditions on this phase, we compare all simulations at the time when their separation first becomes as
    small as $6 \times 10^8\,\mathrm{cm}$. This distance is approximately the distance at which the system described in \citet{pakmor2012a} forms its first hotspot.
    
    Fig.~\ref{fig:rho} shows a density slice at this time. Qualitatively, all four simulations are very similar. In all cases, the primary white dwarf remains mostly unaffected
    and is surrounded by a very thin envelope. There are, however, some quantitative differences in the central density of the secondary white dwarf which is about to
    be disrupted. In particular, for the low resolution it is about $7\times10^6\,\mathrm{g\,cm^{-3}}$, but only about $4\times10^6\,\mathrm{g\,cm^{-3}}$ for the high
    resolution simulations. Although this difference is dynamically not important, it will change the burning products of the secondary white dwarf slightly in the violent
    merger scenario.
        
    At the place where the material from the secondary impacts the primary, a denser region forms. The most interesting region is located between this dense region and
    the surface of the primary white dwarf. Here, material is compressed and heats up enough to start carbon burning which can be interpreted as an indication that a detonation
    can form there. A temperature slice of the binary system is shown in Fig.~\ref{fig:temp} clearly demonstrating the hot region on the surface of the primary white
    dwarf. The plot already shows that the temperature is significantly higher for the high resolution simulations than for the low resolution runs. To quantify this
    difference, we count the number of particles with a temperature larger than $2\times 10^9\,\mathrm{K}$ in all simulations. We find that the runs with high resolution
    have the same number of ``hot'' particles within a factor of two ($81$ particles for the high resolution run with approximate initial conditions and $44$ for the run
    with high resolution and exact initial conditions). The low resolution run with approximate initial conditions contains only $8$ of those ``hot'' particles and the low
    resolution run with exact initial conditions none at all. Note, that with these small numbers of ``hot'' particles none of the simulations are numerically converged.
    As shown in \citet{pakmor2011b}, even with the resolution used in the high resolutions presented here the conditions in the region where the ``hot'' particles emerge
    are actually underestimated compared to even higher resolution.
    
    This result is consistent with other previous simulations with low resolution and exact initial conditions that did not show any carbon burning particles
    \citep{dan2011a,dan2012a} and simulations with both types initial conditions and high resolutions which found ``hot'' particles \citep{pakmor2010a,pakmor2011b,raskin2012a}.
    Therefore, we conclude that the resolution of the simulation is the deciding factor which determines whether hotspots form with particles which started carbon
    burning. The initial conditions, in contrast, do not seem to have a significant effect on this phase of the merger. This result can be understood in terms of timescales.
    The sound crossing time for the secondary white dwarf with a mass of $0.9\,\mathrm{M_\odot}$ is only about $7\,\mathrm{s}$. Therefore, even with approximate
    initial conditions for which the binary system becomes unstable after a bit more than $2$~full orbits, the secondary white dwarf has enough time to adjust its
    structure to the gravitational potential of the primary white dwarf. Moreover, the difference in the total angular momentum of the systems with approximate and exact
    binary initial conditions is too small to significantly affect this phase of the merger.
  
  \section{Summary and conclusion}
    \label{sec:summary}
    In this paper we presented the modifications to the \textsc{Gadget} code that are necessary to apply it to problems of stellar physics. These modifications include
    a general equation of state, a new efficient nuclear reaction network and a mechanism to avoid errors from integrating particles on local timesteps.
    
    We then described a new method to arrange the SPH particles in the initial conditions according to a one-dimensional density and pressure profile as commonly
    given in problems of stellar astrophysics. We
    showed that the configuration we obtain is perfectly stable and retains its radial profiles. We then discussed two different ways of setting up a marginally stable
    binary system from two white dwarfs using approximate and exact \citep{dan2011a} initial conditions.
    
    As an example demonstrating the use of the "Stellar Gadget" code, we investigated the effects of binary initial conditions and resolution on the properties 
    of the merger of a $1.1\,\mathrm{M_\odot}$ and with a $0.9\,\mathrm{M_\odot}$ carbon-oxygen white dwarf \citep{pakmor2012a}. 
    We found that exact initial conditions lead to a significantly more stable binary system, confirming previous
    results by \citet{dan2011a}. However, we also worked out that the type of binary initial conditions has no effect on the properties of the actual 
    merger.
    
    Comparing the simulations in the last binary orbit, when they reach the same separation between primary and secondary white dwarf, we found that
    resolution is the dominating factor. It decides whether carbon burning starts in hotspots or not, regardless of the type of binary initial conditions. We
    stress that merger simulations should be carried out with at least $10^6$ particles to avoid incorrect conclusions due to under-resolved hotspots. 
    
    Unfortunately, our simulations also confirmed again that even for our best-resolved runs the resolution around and at the hotspots is far from sufficient to really
    confirm or rule out the formation of a detonation. Given our results it seems rather unlikely that this will be possible in the near future using an SPH code.
    Thus, the best way to investigate this might be to try to model the accretion stream on a high resolution grid as done for helium white dwarf secondaries in
    \citet{guillochon2010a}.
    
  \section{Acknowledgements}
    We thank Volker Springel and Klaus Dolag for insightful discussions. R.~P.\ gratefully acknowledges financial support of the Klaus Tschira Foundation.
    The work of F.K.R.\ was supported by the Deutsche Forschungsgemeinschaft via the Emmy Noether Program (RO 3676/1-1) and by the ARCHES prize of the
    German Federal Ministry of Education and Research (BMBF).

  \label{lastpage}

\end{document}